%% file: paper.tex
\documentclass[conference]{IEEEtran}
\IEEEoverridecommandlockouts
\usepackage{cite}
\usepackage{amsmath,amssymb,amsfonts}
\usepackage{algorithmic}
\usepackage{graphicx}
\usepackage{subfigure}
\usepackage{textcomp}
\usepackage{hyperref}

\usepackage[table,xcdraw]{xcolor}

\def\BibTeX{{\rm B\kern-.05em{\sc i\kern-.025em b}\kern-.08em
    T\kern-.1667em\lower.7ex\hbox{E}\kern-.125emX}}
    
\usepackage{xcolor}
\usepackage{tikz}
\usepackage[
]{pgf-umlcd}

 \usepackage{amssymb} % \checkmark
 \usepackage{pifont}% http://ctan.org/pkg/pifont
\usepackage{comment}
\usepackage{listings}
\usepackage{fancyvrb}
\usepackage{framed}
\usepackage{color}
\definecolor{light-gray}{gray}{0.95}

\usepackage{todonotes}
\begin{document}

%\title{Deep Learning Scheduler for Vehicular Safety Communication}

\title{Deep Learning-aided Application Scheduler for Vehicular Safety Communication}

%\thanks{This research has been supported by the German Federal Ministry of Education and Research (BMBF) in the project DecADe (16KIS0538).}

\author{Mohammad Irfan Khan$^\dag$, Fran\c cois-Xavier Aubet$^\ddag$, Marc-Oliver Pahl$^\ddag$, J\'er\^ome H\"arri$^\dag$ \\ 
$^\dag$EURECOM, Campus SophiaTech, 450 route des Chappes, 06904 Sophia-Antipolis, France \\
E-mails:~\texttt{\{khanm,haerri\}@eurecom.fr} \\
%$^\ddag$UCL, UK \\
%E-mails:~\texttt{francois.aubet@ucl.ac.uk} \\
$^\ddag$Technical University of Munich, Garching bei Munich, Germany \\
E-mail:~\texttt{\{aubet,pahl\}@net.in.tum.de} \\
}

\maketitle

\begin{abstract}

		802.11p based V2X communication uses stochastic medium access control, which %Therefore it 
		cannot prevent broadcast packet collision, %degrading communication performance 
		in particular during high channel load. Wireless congestion control %Transmit Rate Control 
has been designed to keep the channel load at an optimal point. However, vehicles' lack of precise and granular knowledge about true channel activity, in time and space, % This %lack of control regarding packet scheduling 
		makes it impossible to fully avoid %concurrent transmission and 
        packet collisions.  %with visible and hidden neighbors, 
%		which degrades the performance of broadcast communication for safety V2X services. 
In this paper, we propose a machine learning approach using deep neural network for learning the vehicles'  transmit patterns, and as such predicting future channel activity in space and time. 
%		In this paper, we highlight the challenges of CSMA based medium access for safety V2X communication, and propose a machine learning approach using deep neural network for learning and predicting granular channel activity. %This enables a vehicle to better schedule its packet transmissions and reduce collision. 
		We evaluate the performance of our proposal via simulation considering multiple safety-related V2X services involving heterogeneous transmit patterns.
        % We perform a simulation based evaluation using multiple standardized safety V2X services, %running on each vehicle 
		%having heterogeneous transmit pattern. 
        Our results show that predicting channel activity, and transmitting accordingly, reduces collisions and significantly improves communication performance. %JHNOTE: by XY percent...
        %, both during low and high channel loads.

%802.11p based V2X networks (V2X) provides stochastic medium access control. Therefore it cannot prevent broadcast packet collision, degrading communication performance during high channel load. Although Transmit Rate Control has been designed for V2X broadcast communication, but a vehicle lacks precise knowledge about temporal and spatial channel footprint. This %lack of control regarding packet scheduling 
%makes it impossible to fully avoid concurrent transmission and packet collision with visible and hidden neighbors.

%In this paper, we highlight the challenges of CSMA based medium access of V2X communication and propose a machine learning approach using deep neural network for learning and predicting neighboring transmission. This enables a node to better schedule its transmissions and reduce packet collision. We perform a simulation based evaluation using multiple standardized safety V2X services running %on each vehicle 
%with heterogeneous transmit pattern. Our results show that intelligent scheduling via machine learning reduces collision and significantly improves communication performance, both during low and high channel loads.

\end{abstract}

%\begin{IEEEkeywords}
%\end{IEEEkeywords}

\input{sections/introduction.tex}

\input{sections/problem.tex}
\input{sections/learning.tex}

%\input{sections/scheduling.tex}
\input{sections/evaluation.tex}
\input{sections/related.tex}

\input{sections/conclusion.tex}

\section*{Acknowledgment}
The  authors  acknowledge  the  support  of  the  SCHEIF  project  within  the French-German Academy for the Industry of the Future (www.future-industry.org). EURECOM acknowledges the support of its industrial members, namely, BMW Group, IABG, Monaco Telecom, Orange, SAP and Symantec. This research has been supported by the German Federal Ministry of Economic Affairs and Energy (BMWI) in the project DECENT (0350024A).

\bibliographystyle{IEEEtran}
\bibliography{bibliography}
% \section*{Acknowledgment}
% 
% The preferred spelling of the word ``acknowledgment'' in America is without 
% an ``e'' after the ``g''. Avoid the stilted expression ``one of us (R. B. 
% G.) thanks $\ldots$''. Instead, try ``R. B. G. thanks$\ldots$''. Put sponsor 
% acknowledgments in the unnumbered footnote on the first page.

% \begin{thebibliography}{00}
% \bibitem{b1} G. Eason, B. Noble, and I. N. Sneddon, ``On certain integrals of Lipschitz-Hankel type involving products of Bessel functions,'' Phil. Trans. Roy. Soc. London, vol. A247, pp. 529--551, April 1955.
% \bibitem{b2} J. Clerk Maxwell, A Treatise on Electricity and Magnetism, 3rd ed., vol. 2. Oxford: Clarendon, 1892, pp.68--73.
% \bibitem{b3} I. S. Jacobs and C. P. Bean, ``Fine particles, thin films and exchange anisotropy,'' in Magnetism, vol. III, G. T. Rado and H. Suhl, Eds. New York: Academic, 1963, pp. 271--350.
% \bibitem{b4} K. Elissa, ``Title of paper if known,'' unpublished.
% \bibitem{b5} R. Nicole, ``Title of paper with only first word capitalized,'' J. Name Stand. Abbrev., in press.
% \bibitem{b6} Y. Yorozu, M. Hirano, K. Oka, and Y. Tagawa, ``Electron spectroscopy studies on magneto-optical media and plastic substrate interface,'' IEEE Transl. J. Magn. Japan, vol. 2, pp. 740--741, August 1987 [Digests 9th Annual Conf. Magnetics Japan, p. 301, 1982].
% \bibitem{b7} M. Young, The Technical Writer's Handbook. Mill Valley, CA: University Science, 1989.
% \end{thebibliography}

\end{document}

%% file: sections/introduction.tex
\section{Introduction}
\label{introduction}

Communication between Vehicle to Vehicle (V2V) and Vehicle to Infrastructure (V2I) is being deployed with a goal to improve traffic safety and transport efficiency. Initially a majority of the vehicular safety applications are based on improving a vehicle's awareness of its vicinity by exchanging its  position, speed, heading etc. with neighbors, by periodically broadcasting Cooperative Awareness Message (CAM) or Basic Safety Message (BSM). Further along the road, V2X communication will be used for cooperative driving and navigation, when a variety of messages will be transmitted, as intelligent vehicles will negotiate and coordinate their maneuvers, requiring reliable communication channels.  %The other variant technology is LTE-V2X, which is still in its infancy.
%deployment ready techonology is , while the other nascent technology being LTE V2X
%In the ad-hoc mode of 802.11 based access technology, there is no centralized channel access scheduler. Each node is stochastically granted access using CSMA. Similarly there is no direct coordination or exchange of network control information among nodes. 

Among several potential wireless communication technologies, the technology which is being deployed is called ITS-G5 in Europe and DRSC in the USA, with standardized PHY and MAC layers based on IEEE 802.11p. In the ad-hoc mode of 802.11p, there is no centralized channel access scheduler. Each node is stochastically granted access using CSMA. However, advanced applications such as Autonomous Driving and other safety-V2X services need highly reliable communication, which CSMA based medium access of 802.11p is not capable of providing. As the channel load increases, the communication performance of CSMA degrades rapidly, further affecting the performance of critical V2X services.

%Similarly there is no direct coordination or exchange of network control information among nodes. 

Wireless congestion control has been designed to prevent channel saturation, enabling each node to periodically monitor the channel load and adjusting its transmit rate and power. However, collisions still occur due to the stochastic nature of CSMA and hidden nodes. As safety-V2X services mostly rely on broadcast traffic, packet collisions due to probabilistic channel access or due to hidden terminals cannot be detected nor fully avoided. Yet, what if an intelligent vehicle can precisely anticipate and predict neighboring vehicles' transmission, and accordingly schedule its own transmissions?

%In this paper, 
We address the possibility for a vehicle to learn, predict and transmit channel activities in order to avoid packet collisions. % with neighboring nodes. 
Assuming a vehicle can learn the transmit patters from 1-hop neighbors, it can precisely know the channel activity rather than sensing it. %calculating periodically the aggregate channel load. 
Thus, each node would know much better when to transmit and avoid collisions with its neighbors. Then, if such vehicle %, if a vehicle can learn such transmit patterns and also share 
further shares such predicted channel activity with its 1-hop neighbors, %which would in turn let each vehicle learn the transmit patterns of its 2-hop neighbors, 
it would enable vehicles %2-hop neighbors 
to learn the transmit patterns of hidden nodes. Accordingly, this would let each vehicle not only better schedule its transmissions based on the slots used by its 1-hop neighbors, but also considering those slots sensed `idle' via carrier sense, but actually being occupied by hidden neighbors.

In a static and highly synchronous system, this can be easily optimized by coordinating the transmissions from different nodes. However, V2X communication scenario of safety V2X applications is far from synchronous. The scenario is highly dynamic, with dynamic node mobility, varying neighbor density, fluctuating channel load and external events triggering packet transmissions.  In this regard, machine learning can be an useful tool for such intelligent vehicle to learn and predict its neighbors' transmit patterns.% of its neighbors. 

In this paper, we propose such an approach %to increase a vehicle's knowledge of channel usage 
by learning and predicting neighboring transmissions using Recurrent Neural Networks (RNN) with Long and Short Term Memory (LSTM). Our contributions are threefold: (i) we highlight the challenges of ITS-G5 to sense idle resources in time and space. (ii) we propose a machine learning approach using deep neural network for learning and predicting neighbors’ transmissions. (iii) using simulation based evaluation, we demonstrate that scheduling according to predicted channel activity can significantly reduce packet collision and improve communication performance of safety V2X applications.

The rest of the paper is organized as follows: Section II discusses scheduling and corresponding issues in 802.11p based vehicular networks. Section III presents our approach of increasing a node's awareness and intelligent scheduling via machine learning. Section IV provides performance evaluation results, followed by a brief review of the state of the art in Section V. The conclusion and future work are discussed in Section VI.

%% file: sections/problem.tex
%\section{Channel Congestion Control}

\section{Scheduling in 802.11p Based Vehicular Network}

\label{problem}

\noindent
\textbf{Medium Access:} The medium access of ITS-G5 and DSRC is based on IEEE 802.11 standards, where there is no centralized channel resource scheduler and each node acts decentrally to contend for channel access. It employs a CSMA/CA listen before talk approach, i.e. if the channel is sensed free for a certain time the node transmits directly, otherwise the node chooses a random back-off window, which decreases each time the channel is sensed free. Transmission occurs when the countdown reaches zero. The random back-off value between 0 and CW is chosen to avoid simultaneous channel access by multiple nodes.\\

\noindent
\textbf{Transmit Rate Control:} In CSMA/CA when a unicast packet in not acknowledged, the contention window is enlarged. This reduces channel congestion by distributing the transmission attempts over longer period. However, safety related vehicular communications involve packet broadcast without acknowledgment, so this contention window enlargement is not possible. To counter this problem, on top of CSMA, there is additional flow control to limit the transmit rate of each node and reduce channel congestion. This mechanism is also known as Decentralized Congestion Control, or DCC in European Standards.

%Each node periodically probes the channel and calculates the channel load as the ratio of the number of slots busy over total slots in an observation window of 100ms. If the channel load exceeds a threshold, the node limits its transmit rate by delaying or dropping packets via queuing and flow control.

%activity and calculates the channel busy ratio (CBR) if the ratio of the number of samples found busy over the total samples during the observation window of 100ms, according to:

%\begin{equation}
%CBR_{measured} =  \frac{Samples_{busy}}{Samples_{per\_cycle}} 
%\end{equation}

%Based on the CBR, the node adapts its Transmit Rate by delaying or dropping packets via flow control.

%\noindent
%\textbf{Packet Generation Control:} In addition to Flow control at the MAC layer, the generation of periodically broadcast CAMs depends on node dynamics and can be delayed by packet generation control at the application layer to respect the node's duty cycle limit during channel congestion.

%However, the packet scheduling via these three steps is not sufficiently optimized based on the available temporal and spatial channel usage information and has issues as explained in the next subsection.

\subsection{Issues with existing Approach}
\textbf{Stochastic Medium Access:} CSMA attempts to minimize concurrent channel access by several nodes using a random back-off window, usually of size between 0 to 15 slots. However, it is still probable for two nodes to obtain the same back-off window or same remaining back-off. Identical back-off results in simultaneous transmissions and collision.\\
%As the number of contending node increases, so does the probability of this type of collision. 
%Moreover, if multiple nodes simultaneously perform a carrier sense and find the medium free at the first attempt, they transmit without random back-off, resulting in collision.\\  %, in spite of nodes controlling their transmit rates via DCC.\\ 

\noindent
\textbf{Lack of Spatial Resource Reuse:}
The presence of hidden nodes beyond the Carrier Sense range cannot be detected via Carrier Sense. This results in packet collision and deteriorates the communication performance significantly as the node density increases. CSMA does not employ information of spatial channel usage, beyond the range of Carrier Sense. For example, if hidden nodes could transmit during different time slots, it could mitigate the problem of hidden node collision.\\

\noindent
\textbf{Lack of a notion of Scheduling:}
The goal of CSMA is to stochastically attribute channel access to avoid concurrent transmissions by several nodes. Additionally during high channel load, transmit rate control limits the transmit rate of each node to prevent channel saturation. However, CSMA or transmit rate control do not aim to schedule or uniformly distribute the transmissions of the nodes along the time axis in a coordinated manner.\\

%During a 100ms cycle,

\noindent
\textbf{Channel Load calculation Granularity:}  Along the time axis, there can be periods of higher channel footprint during transmission bursts, when more nodes will contend for channel access. Although most transmissions are periodic or quasi-periodic during initial vehicular network deployment, in future some vehicles will have more advanced capabilities. Those vehicles will transmit multiple packets with different transmit patterns, which will result in variations of channel footprint. This is impossible to observe by the present mechanism of channel load measurement. In the standards, the channel load is filtered and calculated once at the end of a 100ms window, while the vehicle is unaware of the channel activity during the rest 99ms. This will degrade communication performance for future deployment scenario, involving heterogeneous and multiple safety applications per vehicle.

%% file: sections/learning.tex
\section{Intelligent Scheduling via Machine Learning }
\label{learning}

\begin{figure}[t!]
	\centering
	\includegraphics[width=1\columnwidth]{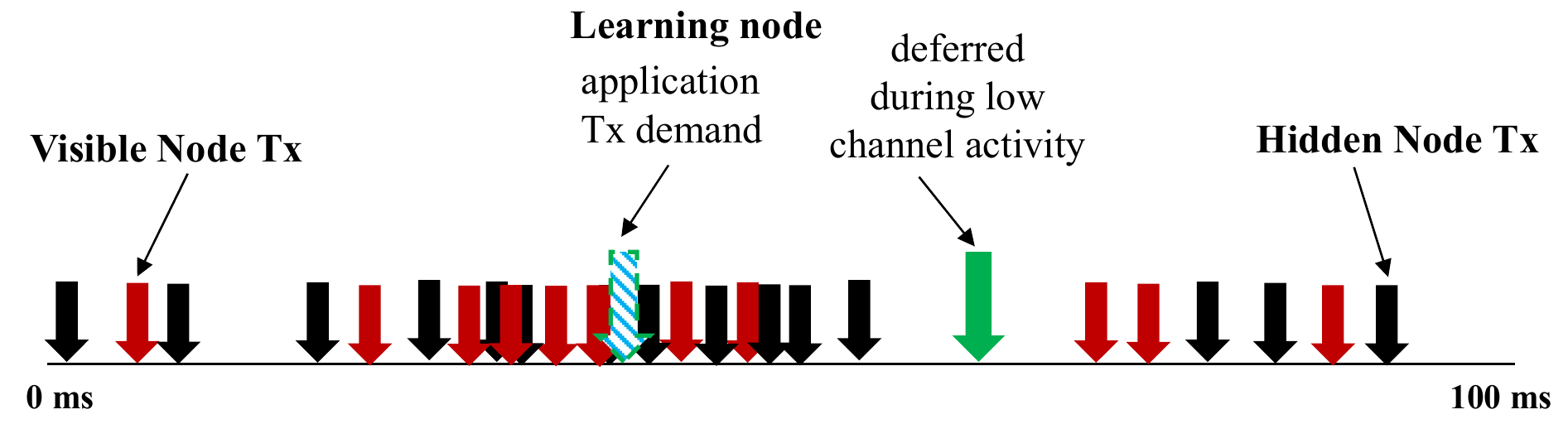}
	\caption{Transmission deferred to period of low channel activity}
	\label{fig:Gap} 
\end{figure}

\begin{comment}
\begin{figure}[t!]
	\centering
	\includegraphics[width=1\columnwidth]{images/No_Gap.pdf}
	\caption{High Channel Footprint}
	\label{fig:NoGap} 
\end{figure}
\end{comment}

%\subsection{Proposed Solution}
In this section we present a learning node, which learns the channel activity during an observation window of 100ms and predicts neighbors' packet transmissions, packet size, type and the channel footprint for the next few windows of 100ms. %100ms and next few cycles. 
The goal is to use the learned pattern of neighbors' packets and schedule its own packets, depending on the application deadline, during periods of low or no channel activity, as shown schematically in Figure \ref{fig:Gap}. 

The figure shows a typical prediction pattern of a learning node, predicting the time instances when neighbors will transmit during the next 100ms. The dotted arrow indicates that an application of the learning node needs to generate a packet at a certain point. However, according to the prediction pattern, a period of low channel footprint will be available in the current prediction window. Consequently, the application defers the packet generation and eventually generates and transmits the packet during a period of lower channel activity. 

The tolerated delay of deferring a packet depends on the application requirement. The goal is to decrease the probability of concurrent transmissions, and avoid interfering with visible and hidden neighbors, while remaining within the packet transmission deadline requirement of the application.

The learning node monitors all received packets from visible neighbors and uses the packet reception history to predict its neighbors' future transmissions. Furthermore, each node piggybacks the packet reception pattern of its own neighbors inside the packets it transmits. Thanks to this piggybacking, the awareness of the learning node is extended and it becomes aware of the transmit patterns of hidden nodes as well. %two hops away. %which are outside its carrier sense range, but close enough to interfere with its transmissions to its one hop neighbors. 

\begin{figure}[t!]
	\centering
	\includegraphics[width=1\columnwidth]{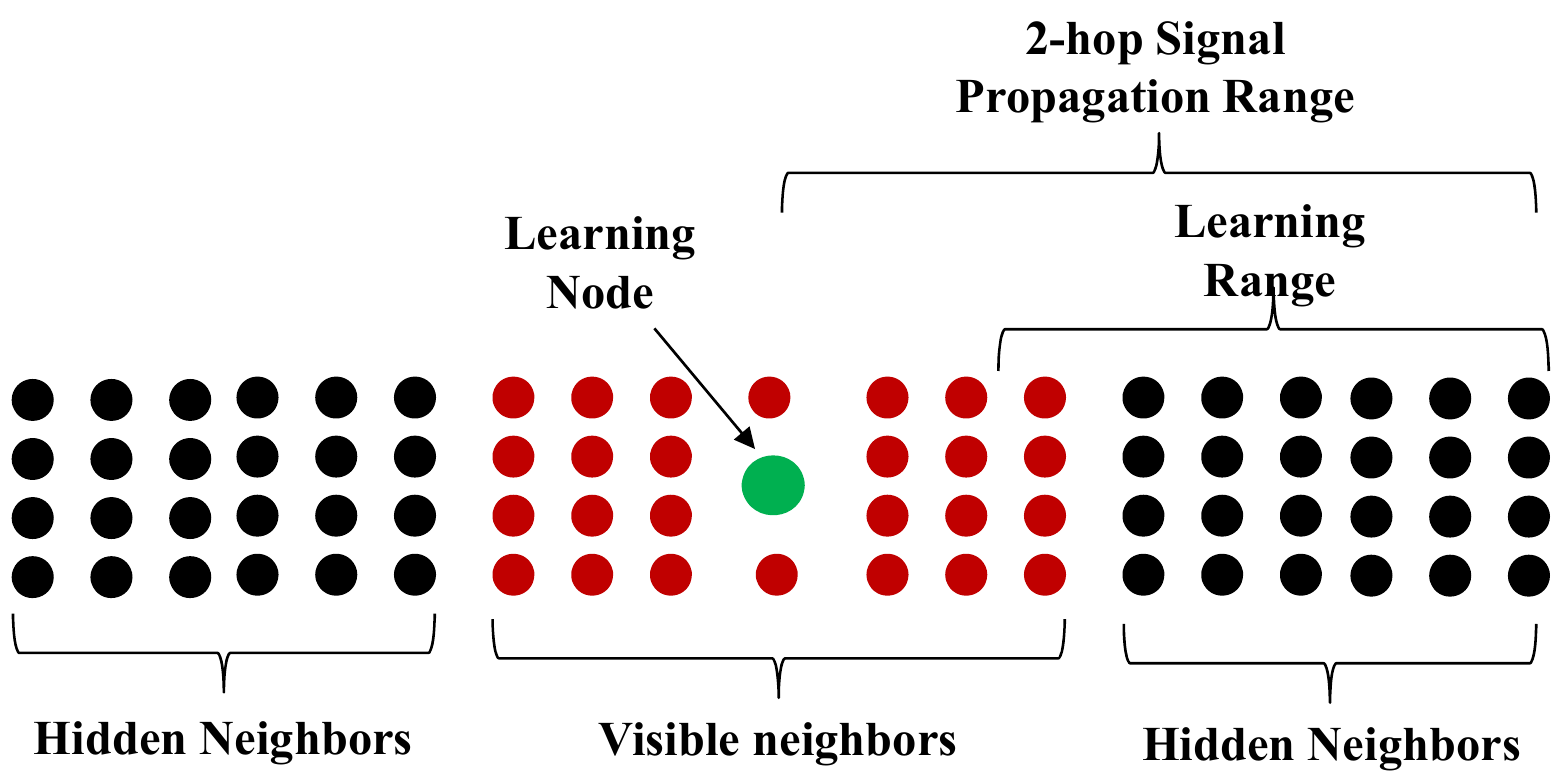}
	\caption{Learning Distance of Intelligent Node}
	\label{fig:LearningDistance} 
\end{figure}

Nevertheless, the number of neighbors a learning node can keep track of and predict their transmissions is limited. If a leaning node has to keep track of a large number of neighbors, such as in a scenario of high vehicle density, then it is difficult to find vacant windows to schedule is own transmissions. %, as shown in Figure \ref{fig:NoGap}. The figure is a schematic prediction pattern, when a learning node has to keep track of a high number of neighbors. 

The set of 1-hop visible and 2-hop hidden neighbors that a learning node can keep track of has to be chosen optimally. %It has to prioritize hidden nodes first and then visible nodes, as shown in 
Figure \ref{fig:LearningDistance} shows a schematic scenario of learning during a high node density. In the figure, the green point indicates the learning node, the red points indicate the nodes visible to the learning node and the black points are the hidden nodes. In such a scenario, the learning node prioritizes learning and predicting the transmit patterns of hidden nodes 2-hops away. As detailed in the next section, collisions due to hidden nodes play a more significant role in degrading communication performance, while potential collisions due to visible nodes are largely prevented by CSMA/CA. %This is discussed further in the next Section. 

%Similarly, learning neighbor's transmit patterns and predicting microscopic channel footprint for the next several windows permits a node to better schedule packets based on priority and space out its transmission in time in order to avoid spikes in channel footprint. For example, during windows of high channel footprint, it can schedule only delay critical packets during those windows and defer packets during 100ms windows of relatively lower footprint. Thus keeping track of neighbors transmission can add a lot of optimization to a node's transmission instead of simply relying on transmit rate control with CSMA.

%\todo[inline]{Francois writes this}

%Time series prediction has been done extensively for the problem of vehicle traffic flow prediction. Over the years, many different algorithms have been proposed, the earliest being AutoRegressive Integrated Moving Average (ARIMA). \cite{lv2015traffic} Followed by prediction methods using different types Artificial Neural Networks (ANN). The authors of \cite{tian2015predicting} have proposed an approach of time series prediction using RNN with LSTM. %Their results show that this approach achieves higher accuracy and generalizes better than the other methods.

\subsection{Machine Learning for Predicting Neighbors' Transmissions}

For predicting vehicular message transmissions, we use time-series prediction using RNN with LSTM. There are many algorithms for predicting sequential data, the earliest algorithm being AutoRegressive Integrated Moving Average (ARIMA). For most use cases, ARIMA or Hidden Markov Models (HMM) have become deprecated and have been replaced by RNN, for reasons outlined in \cite{lipton2015critical}. %RNN. even though more complex than ARIMA, allow a better performance. 

% Other less detailed explaination:
The algorithms used to train HMM and vanilla RNN struggle to deal with many different inputs and to capture long term dependencies. %, as it increases the size of the state space and the transition matrix grows according to the square of the state space.
For the use case of predicting messages of neighboring vehicles, the consequence would be that the influence of older messages on the current prediction would be ignored.
LSTMs were designed to overcome this issue as discussed in \cite{hochreiter1997long}. For these reasons, we decided to use RNN with LSTM to predict messages from vehicles.

% should I explain more here?

%\todo[inline]{should i put more explaination here?}

%- Explanation of Machine learning technique we use:  Recurrent Neural Network, Long Short Term Memory
%- What are the features used to be predicted, and why this machine learning is suitable for this type of data

\subsection{Design of the Predictor}

\begin{figure}[ht!]
	\centering
	\includegraphics[width=0.95\columnwidth]{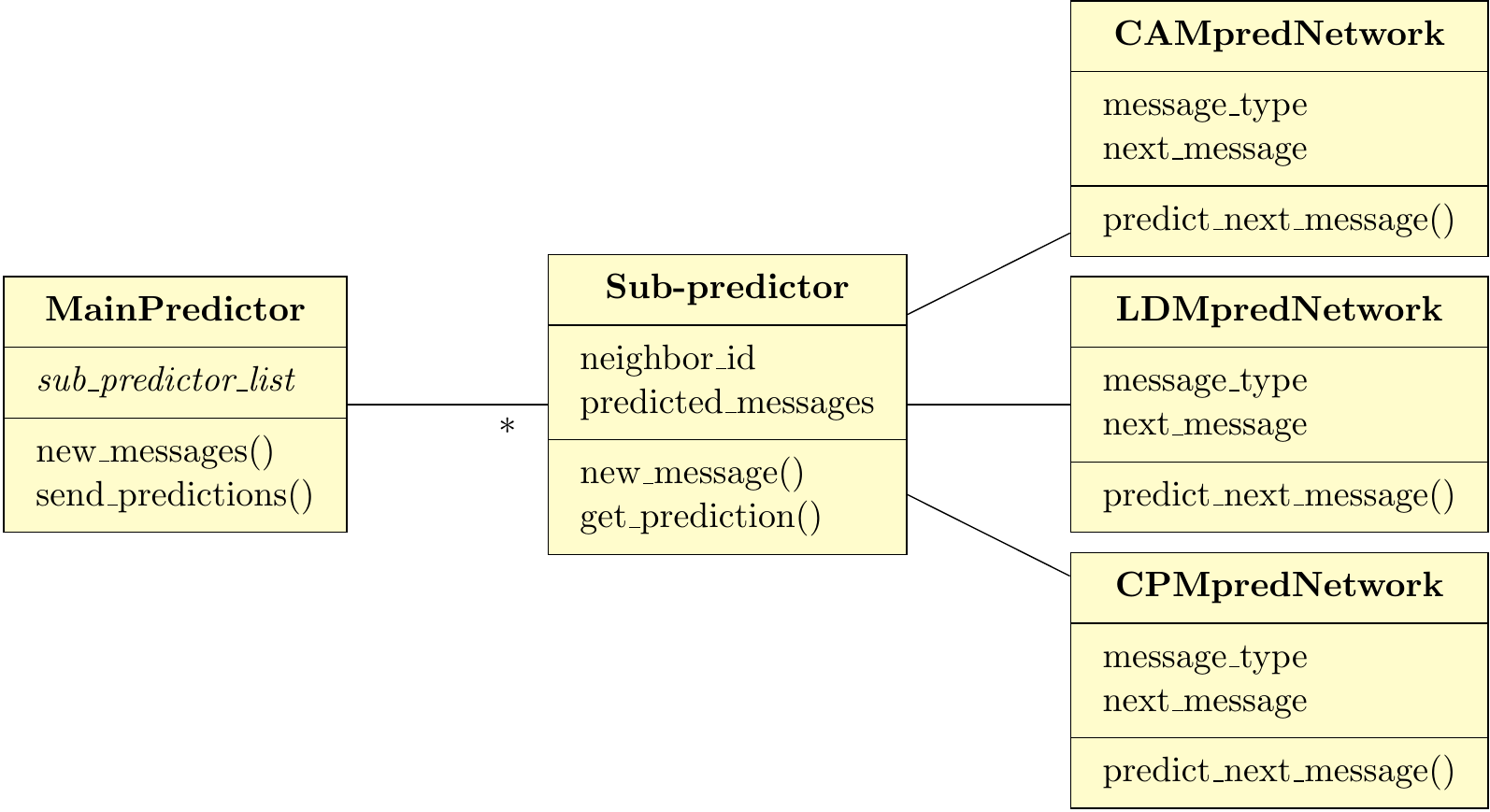}
	\caption{Predictor Architecture}
	\label{fig:predictor} 
\end{figure}

%Explain the prediction architecture using block diagram
%explain the architecture of the predictor program (main, box)
%explain why it makes sense to "divide and conquer" the prediction this way
%explain how the prediction is gathered and what is predicted

In order to predict messages from neighboring vehicles, the learning vehicle uses a \textit{divide and conquer} approach. It maintains a sub-predictor instance for each neighbor, and predicts the neighbor's future packets based on the previous ones. The predictor is trained off-line, using the typical communication pattern of a vehicle. %Different instances of this predictor are used by the learning node for predicting neighbors' packets. 
The sub-predictor uses one RNN for each type of packet. 

%Instead of maintaining a predictor for each neighbor another approach is to train and maintain only one predictor that would predict the messages of all neighbors. However, even given a huge number of training examples, there is no guaranty that this method would be able to predict messages of unseen scenarios. Moreover, the packets of the different neighbors are independent of each other, this means that they can be considered separately.
%However, this method is more complicated, requiring a huge number of training examples, so has not been adopted in this work.

The organization of the prediction program can be seen in Figure \ref{fig:predictor}. The main predictor keeps an active instance of the sub-predictor for each of the current neighbors. The sub-predictor handles all the packets received from a particular neighbor. It uses them to predict the next packet of each type from that neighbor.
When a new packet is received by the sub-predictor, it pre-processes the packet to obtain the information used by the neural network and then feeds it to the corresponding neural network. 

Periodically, every 100 milliseconds, the learning node inquires the predictor for the predicted packets for the next 100ms. The main predictor iterates through all the active instances of the sub-predictors to fetch packet predictions, and returns a complete list of future packet transmissions and the packet air time. After a time-to-live, if no more packets are received from a neighbor, the corresponding sub-predictor instance is deleted. This means that the neighbor has moved out of the learning node's communication range and is no longer relevant. %in an other direction so there is no need to waste memory.

\subsection{Features selection and preparation}

%As explained earlier, the learning node predicts three types of packets coming from each neighbor, i.e. CAMs (motion-event triggered and periodic), event triggered CPM bursts and periodic LDMs. These packets are further detailed in the next Section.

The learning node predicts three types of packets transmitted by each neighbor, i.e. CAMs (motion-event triggered and periodic), event triggered bursts of Cooperative Perception Message (CPM) and periodic exchange of High Definition Maps between vehicles, using a message called Local Dynamic Map (LDM). These packets are further explained in the next section.

%As shown in Figure \ref{fig:predictor}, we train and use a separate neural network for each type of packet, i.e. future LDM packets are only predicted by the LDMpredNetwork.
For each type of packet, a separate neural network is used. Each neural network receives as input the time interval between currently received packet and the previous packet of the same type from a particular neighbor. 
Conceptually, this means that the interval to the next packet is predicted using the interval between the two previous packets.
%Using the duration between two packets allows us to have measurements that are independent from the date and time.

CAMs are triggered by a change in a vehicle's speed, direction or position, and the values of speed, direction and position of the CAM sender are contained inside the CAM. These values of vehicle dynamics and their gradients are also fed into the neural network. All these input features are normalized before being fed to the RNN. We use feature scaling to map the values between -1 and 1.

%\subsection{Training}
%\todo[inline]{Explain about the dataset here}

%% file: sections/evaluation.tex
\section{Evaluation}
\label{evaluation}

%\todo[inline]{Irfan writes this}

\begin{table}[b!]
	
	\centering
	\caption{Simulation Parameters}
	\label{params}

\begin{tabular}{|l|l|}
	\hline
	\multicolumn{1}{|c|}{\textbf{Parameter}} & \multicolumn{1}{c|}{\textbf{Value}}                                                                                                                          \\ \hline
	Transmit Rate                            & \begin{tabular}[c]{@{}l@{}}CAM: 10 {[}Hz{]} \& Triggered\\ CPM: 5 {[}Hz{]}, LDM: 1 {[}Hz{]}\end{tabular}                                                     \\ \hline
	Transmit Power                           & 20 dBm                                                                                                                                                       \\ \hline
	Packet Size                              & \begin{tabular}[c]{@{}l@{}}CAM: 300 Bytes, CPM 500 Bytes\\ LDM: 750 Bytes\end{tabular}                                                                       \\ \hline
	EDCA Packet Priority                     & \begin{tabular}[c]{@{}l@{}}CAM: Best Effort, \\ CPM \& LDM:  Background\end{tabular}                                                                         \\ \hline
	DataRate                                 & 6 Mbps                                                                                                                                                       \\ \hline
	Mobility                                 & \begin{tabular}[c]{@{}l@{}}3 by 3 lane 10 km highway\\ Speed 20 to 45 {[}m/s{]}\\ Gauss Markov, Memory level 0.95\\ Sampling period 0.1 {[}s{]}\end{tabular} \\ \hline
	Node Density                             & 50 vehicles/lane/km                                                                                                                                          \\ \hline
	PHY and MAC                              & \begin{tabular}[c]{@{}l@{}}ITS-G5 802.11p in 5.9 GHz\\ (10 MHz Control Channel)\end{tabular}                                                                 \\ \hline
	Attenuation                              & Log Distance Path Loss                                                                                                                                       \\ \hline
	Preamble Detection Threshold             & - 95 dBm                                                                                                                                                     \\ \hline
	Neural Network                           & \begin{tabular}[c]{@{}l@{}}4 layers: 40, 50, 60 neurons \&\\ LSTM unit layer\end{tabular}                                                                    \\ \hline
	Training                                 & \begin{tabular}[c]{@{}l@{}}Off-line, ADAM algorithm\\ Stochastic gradient descent\end{tabular}                                                               \\ \hline
	Performance Indicators                   & \begin{tabular}[c]{@{}l@{}}Packet Reception Ratio\\ 50 runs, 95\% Confidence Interval\end{tabular}                                                                  \\ \hline
\end{tabular}
\end{table}

We perform a simulation based evaluation to demonstrate the communication performance improvement achieved by learning and predicting neighbors' transmissions, and transmitting during periods of low channel usage. %Our goal is to demonstrate that 802.11p MAC protocol has performance issues, and it can be made more robust by taking intelligent decisions about packet generation at the higher layers. 

We analyze the effectiveness of our machine learning method in reducing collision with the transmissions of visible nodes within 1-hop distance, and hidden nodes beyond the range of carrier sense, within 2-hop range. The packet reception ratio by the neighbors of the learning node for various distances is the primary metric of performance evaluation.

%and we also consider the Inter Reception Time (IRT) to give an idea of the performance benefits for applications due to the learning mechanism.

A 10km long dense highway scenario is used, consisting of 50 vehicles/lane/km and 3 lanes in each direction. Vehicles move at speeds between 20 to 45 m/s, following a Gauss-Markov mobility model. The simulator used is called iTETRIS \cite{itetris}, which has a full ITS-G5 protocol stack implemented on top of NS-3. %A transmit power of 20dBm is used which gives a signal propagation range of 250 meters, resulting around 140 one hop neighbors which are visible, and similarly 140 more neighbors which are hidden. The nodes at the edges are not considered to ignore boundary affect. In order to zoom in on packet loss solely due to collision from visible and hidden neighbors, the attenuation is log-distance and the random aspects of signal propagation such as fading and shadowing have not been considered in this paper and are left for future work.

%The signal propagation is Line of Sight (LOS) and, the random aspects of signal propagation such as fading, shadowing, NLOS propagation etc. are not considered in this paper and left for future work. 

We consider 3 types of packets i) CAM (periodic 10Hz and motion triggered), ii) CPM (bursts) and LDM (periodic). In European standard ETSI EN 302 637-2 \cite{etsi302637}, CAMs are generated as a function of change in vehicle dynamics, either 4m variation in position or 4 degree change in heading or 0.5m/s difference in speed. %These features of motion are used by a learning node to learn the pattern and predict the next packet from the emitter. 
We also consider CAM at 10Hz, as Basic Safety Message (SAE J2735 \cite{saeJ2735})  which is the equivalent in USA are emitted at 10Hz. The CAM size we use is 300 Bytes.

%The CPM messages are currently being standardized and the triggering conditions has not been finalized and we followed the message rate limit of 5Hz as in the latest version of the standards.  Nevertheless, 
CPM is being standardized in ETSI TS 103 324 \cite{CPM}, and is triggered upon detection of vulnerable road objects, which in our simulation is triggered randomly, and 5 messages are emitted within 500ms. Unlike CAM, CPM is not mandatory and only vehicles with appropriate object detection capability will generate CPM. Thus 50\% of the nodes in our simulation emit CPM with message size of 500 Bytes.

Lastly LDM as described in ETSI TR 102 863 \cite{LDM} are messages intended to exchange HD maps of cars, and are emitted periodically at 1 Hz in our simulation settings with message size of 750 Bytes. The nodes start transmission following an uniform random distribution and there is a small jitter of 500$\mu$s during transmission of each packet.  The results are averaged over 50 simulation runs with 95\% Confidence Interval.

%For machine learning and prediction, the LSTM with RNN have been implemented in tensorflow. The neural network consists of 4 hidden layers, with 40, 50 and 60 neurons and a LSTM unit layer. The training is done off-line using the ADAM algorithm, using stochastic gradient descend with a batch of size 1. Table \ref{params} summarizes the main simulation parameters.

For machine learning and prediction, the LSTM with RNN have been implemented in tensorflow. The neural network consists of 4 hidden layers, with 40, 50 and 60 neurons and a LSTM unit layer. This neural network size is a trade-off for this use case, which is large enough to capture the complexity of the data, and small enough to be trained efficiently. 

The training is done using the ADAM Optimizer, with stochastic gradient descent. The batch size is 1 in order to capture the time dependencies between the packets. The training is done off-line using packets logged during simulation runs on highway scenarios. The prediction is done on-line during the run time as the learning node receives transmissions from its neighbors. Table \ref{params} summarizes the main simulation parameters.

 %In order to adapt to changes in the standards that would cause the transmission patterns to change, batch retraining could be used.
%Nevertheless, in real vehicular communication scenarios, the number of services are fixed during a vehicle's runtime. They are all activated when the engine is started, and no new service pops up during the vehicle's runtime. 

%Lastly, one key point to be noted for vehicular networks is that 

\begin{table}[]
	
	\centering
	\caption{Average Channel Load for different transmit Patterns}
	\label{CLtable}	
	
	\begin{tabular}{|l|c|}
		\hline
		\rowcolor[HTML]{9B9B9B} 
		\textbf{Transmit Pattern}           & \multicolumn{1}{l|}{\cellcolor[HTML]{9B9B9B}\textbf{Average Channel Load}} \\ \hline
		\textbf{10 Hz CAM}                  & \textbf{65.35 \%}                                                             \\ \hline
		\textbf{Triggered CAM Higher Speed} & \textbf{50.74 \%}                                                             \\ \hline
		\textbf{Triggered Lower Speed}      & \textbf{35.47 \%}                                                             \\ \hline
		\textbf{CAM + CPM}                  & \textbf{52.10 \%}                                                             \\ \hline
		\textbf{CAM + CPM + LDM}            & \textbf{66.90 \%}                                                             \\ \hline
	\end{tabular}
\end{table}

%Then explain later the different combinations of simulation runs.....
%Moreover, even if a new CAM can be emitted upon the necessary variation in vehicle dynamics, however the condition to check the variation happens at periodic instances, known as \textit{TCAMcheck} in EN 302 637-2, which can be maximum 100ms. This period of \textit{TCAMcheck} discretized even further the time instances of the next CAM emission, making prediction even easier. %The periodic packets are the easiest to predict using a simple time series. Lastly, the event triggered bursts are predicted via.

%We evaluate a heterogeneous scenario, where all vehicles have at least the mandatory CAM service running, while some vehicles additionally the periodic packets and few vehicles emit all the 3 packets. 

%Give the channel load values here.... as a table  Make life easy for the reader

%Change the name from 0 to 250 to one hop and 2 hop neighbors.

%Talk also about contiguous slots and the middle of it.
%Explain all the aspects Jerome talked about thursday night....

\begin{figure}[b!]
	\centering
	\includegraphics[width=1\columnwidth]{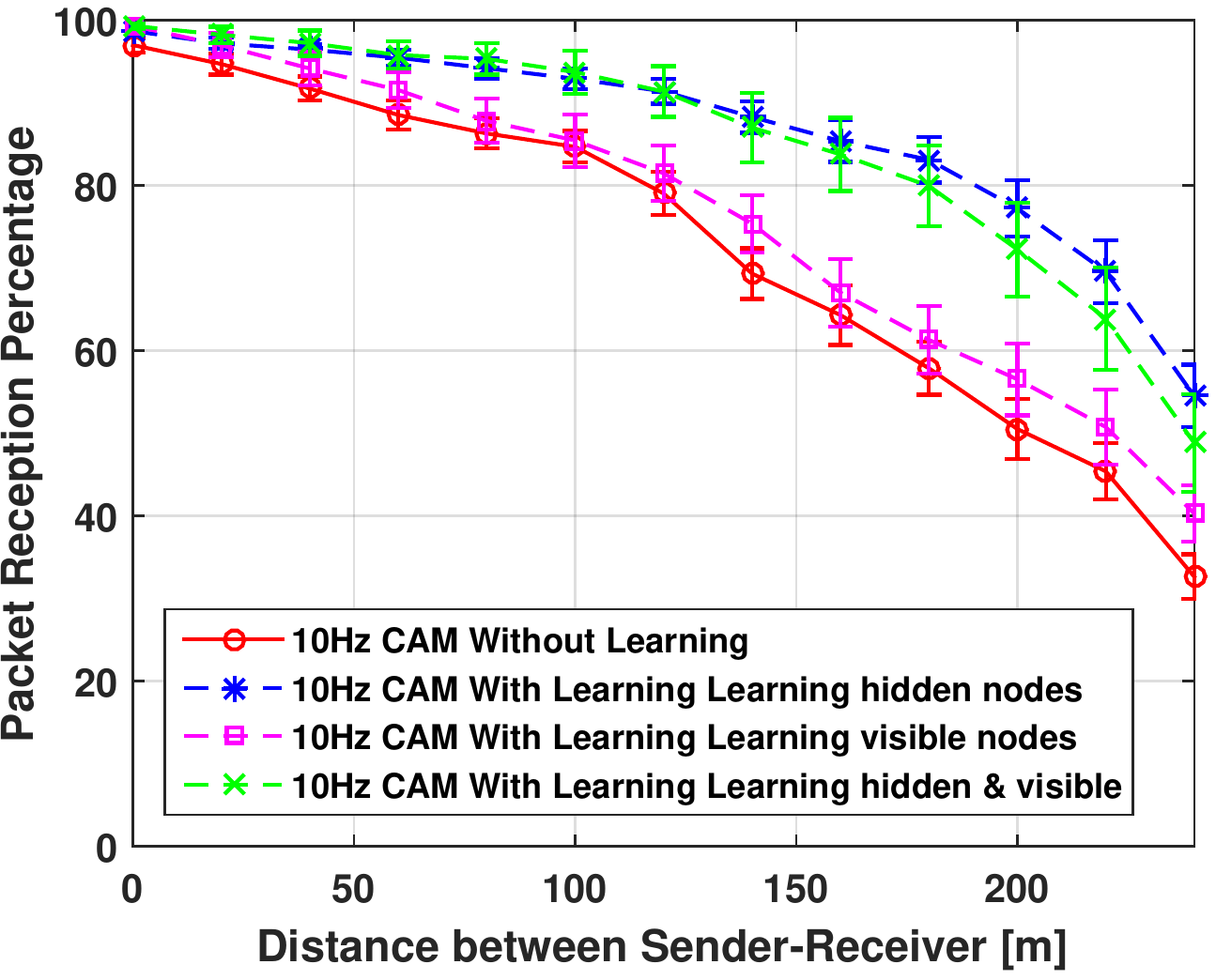}
	\caption{Packet Reception Ratio of 10 Hz Periodic CAMs}
	\label{fig:10HzCAM}
\end{figure}

\begin{figure}[b!]
	\centering
	\includegraphics[width=1\columnwidth]{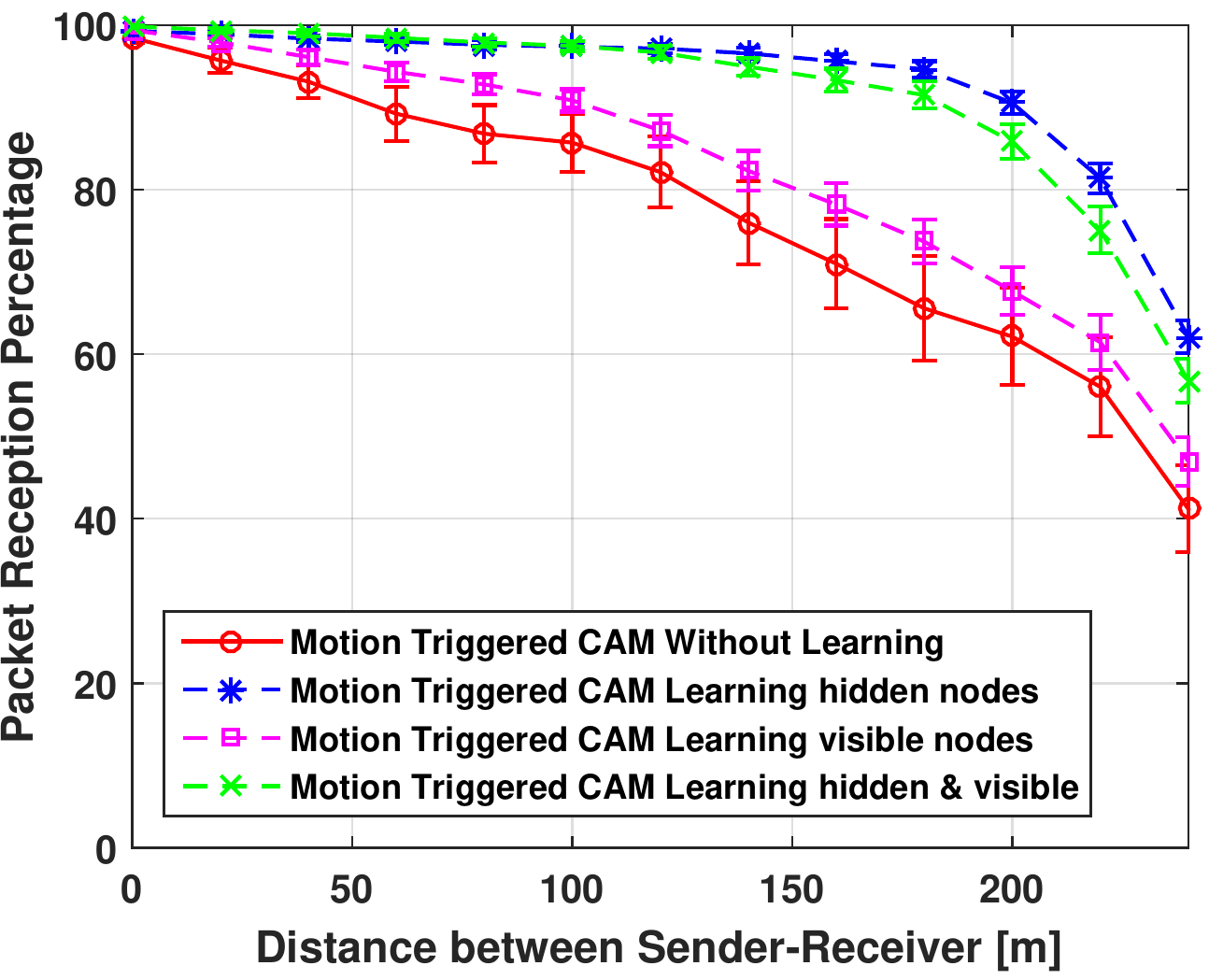}
	\caption{Packet Reception Ratio of Triggered CAMs for vehicle speed of 35-45 m/s}
	\label{fig:TrigCAMHi}
\end{figure}

\begin{figure}[b!]
	\centering
	\includegraphics[width=1\columnwidth]{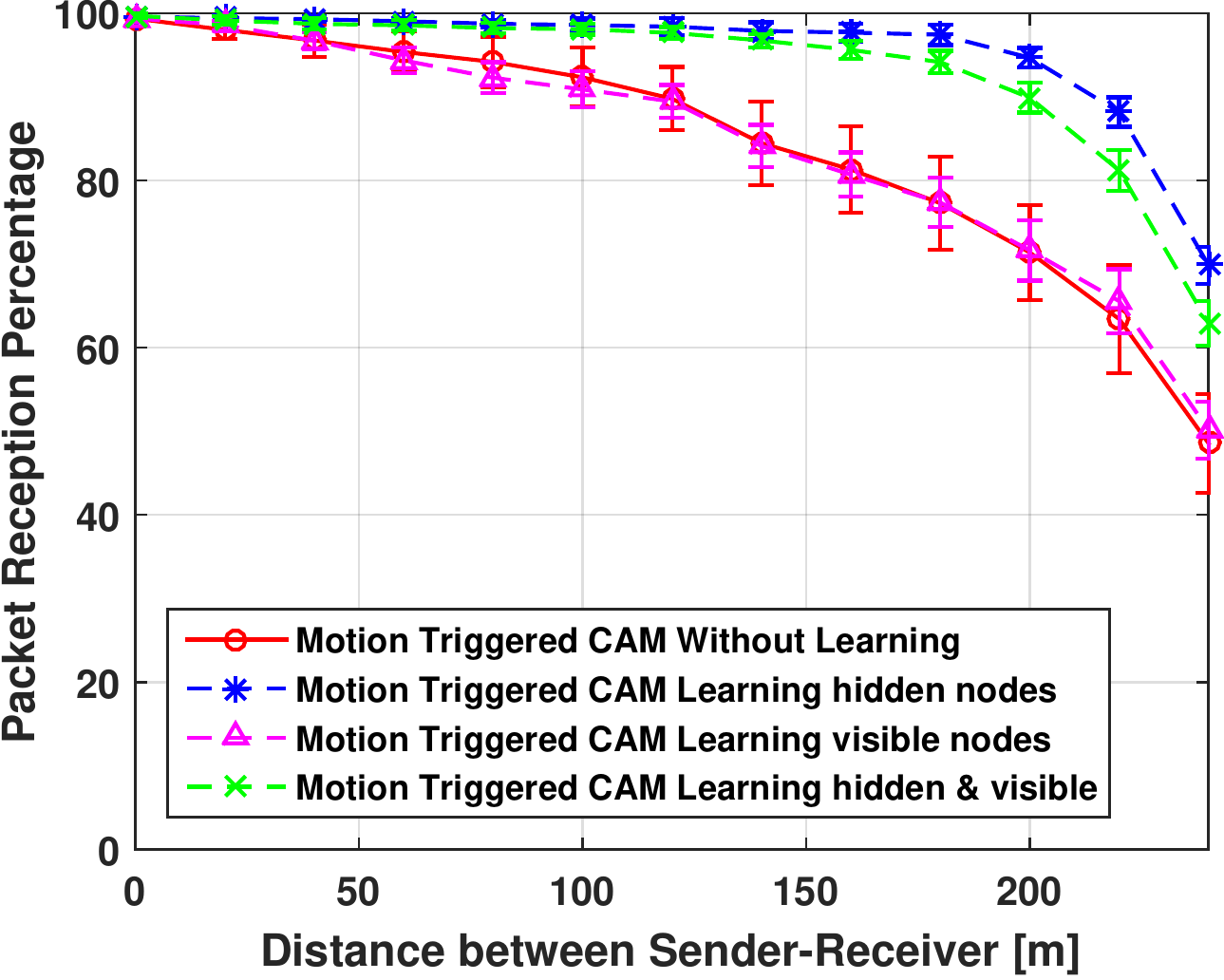}
	\caption{Packet Reception Ratio of Triggered CAMs for vehicle speed of 20-30 m/s}
	\label{fig:TrigCAMLo}
\end{figure}

\begin{figure}[b!]
	\centering
	\includegraphics[width=1\columnwidth]{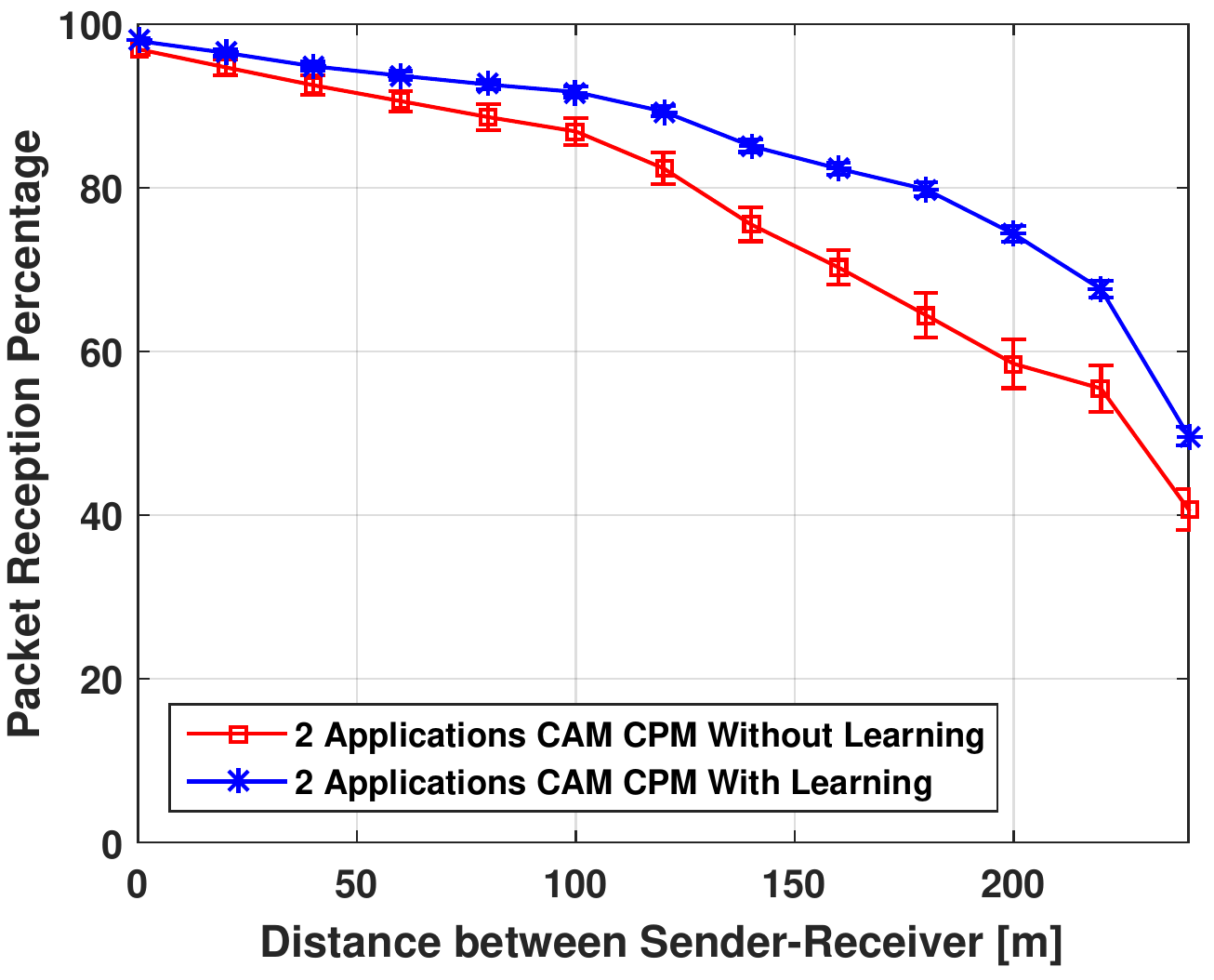}
	\caption{Packet Reception Ratio of Two Applications CAM and CPM}
	\label{fig:TwoPacket}
\end{figure}

\begin{figure}[ht!]
	\centering
	\includegraphics[width=1\columnwidth]{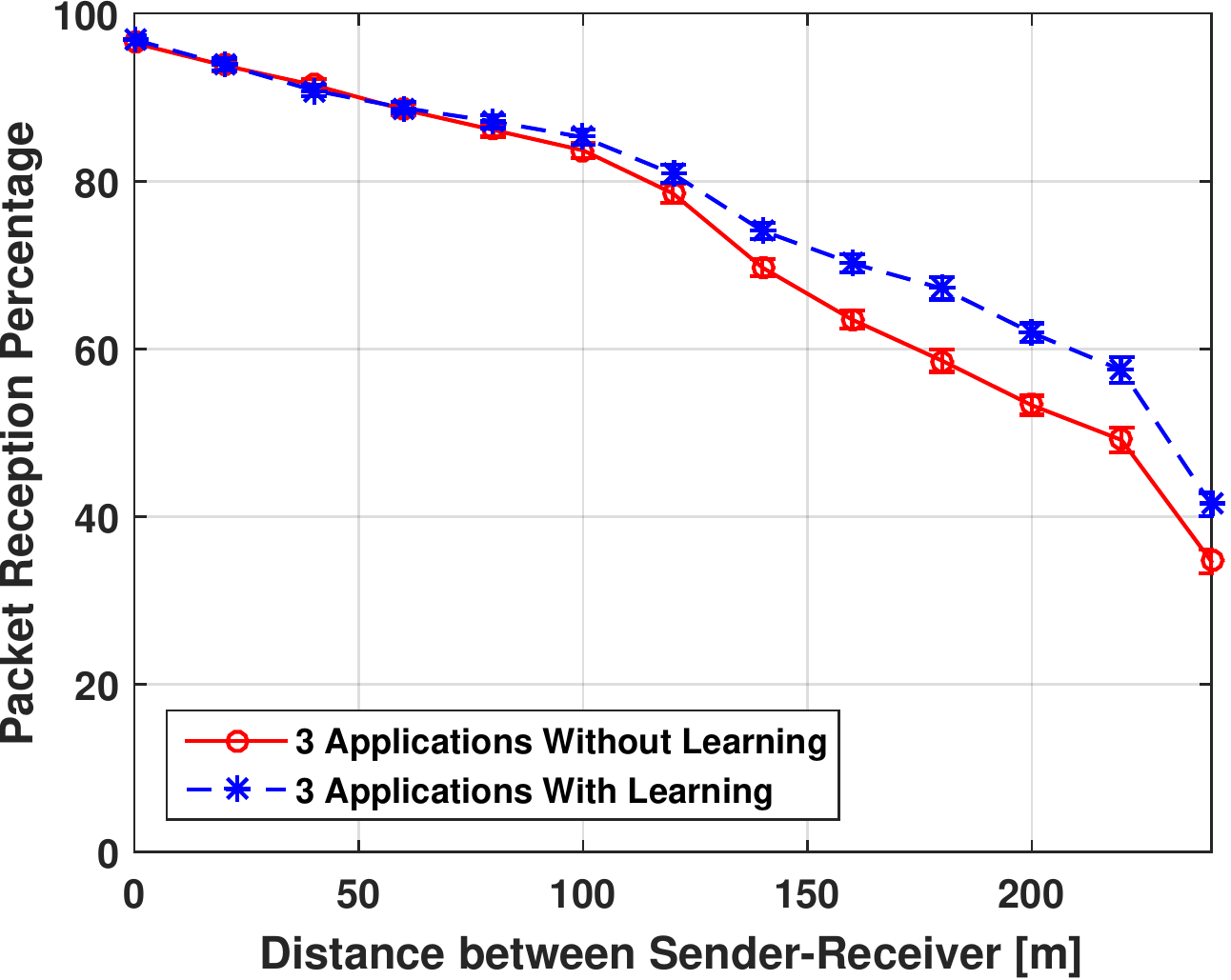}
	\caption{Packet Reception Ratio of Three Applications CAM, CPM, LDM}
	\label{fig:MultiPacket}
\end{figure}

Figure \ref{fig:10HzCAM} shows the packet reception ratio (PRR) on the y-axis by the neighbors of the learning node when vehicles emit 10Hz CAMs, producing an average channel load of 65.35\% as shown in Table \ref{CLtable}. The x-axis corresponds to the distance between the learning node and the receiving nodes. %The packet reception rate decreases with distance as the signal power gets attenuated. %Similarly, if the transmission of the sender interferes with transmission of hidden nodes, the receiver decodes the stronger signal. %, as phenomena known as capture effect.  

The case with no learning performs worse compared to when a node transmits according to predicted transmissions of its visible and hidden neighbors. %As can be seen in the figure, the case with no learning has the lowest PRR for all distances between the sender and receiver.
%to avoid synchronous transmission. 
The reception performance is improved a bit by predicting and avoiding concurrent transmissions with 1-hop visible neighbors. However, the performance improvement is the highest, when the learning node predicts the transmissions of hidden nodes. This indicates that collisions with hidden nodes play a more significant role in performance degradation, than visible nodes. 

Nevertheless, when the learning node predicts the transmissions of both 1-hop visible and 2-hop hidden nodes, the performance reduces a bit than the case with learning simply hidden nodes. In the simulations, within a distance of 2-hop signal propagation range, there are approximately 280 nodes in total across 500m in both directions. In this case the learning node cannot find enough vacant periods to schedule its own transmissions. Nevertheless, transmitting to avoid concurrent transmissions with hidden nodes, produces an improvement of 10\% and 25\% PRR, at distances of 100m and 200m respectively.

Figure \ref{fig:TrigCAMHi} shows the PRR when CAMs are triggered according to vehicle dynamics. Compared to 10Hz transmission, the PRR is higher, as a velocity between 35 to 45 m/s triggers CAMs between 5 and 10Hz creating a lower channel load of 50.74\% compared to 65.35\% channel load produced by the earlier scenario of 10Hz periodic CAMs. A lower channel load results lesser collisions, giving a better PRR. 
The trend is similar, i.e. learning only hidden nodes' transmissions performs the best, followed by learning both hidden and visible nodes, then learning only visible nodes. Lastly the case of no learning performs the worst.

This trend continues when the channel load gets even lower at 35.47\% for a velocity of 20-30m/s as shown in Figure \ref{fig:TrigCAMLo}. However, with a low channel load of 35.47\%, the collision with visible nodes is almost negligible, therefore learning the transmissions of only visible nodes provides no improvement. At a distance of 200m, improvement in PRR due to learning is around 30\% and 35\% for channel loads of 50.74\% and 35.47\% respectively, for CAMs triggered at higher and lower speeds.

%Give numbers to performance improvements and etc.
%As the channel load decreases, the proportion of collision with visible neighbors decreases thereby learning only one hop neighbors doesn't improve much. There are several factors:  channel load, Why learning of visible has more effect with 10Hz, compared to others ?

In addition to single CAM application, we analyze the packet reception performance when 50\% of the nodes emit Cooperative Perception Messages (CPM) to broadcast their sensor information. The PRR is shown in Figure \ref{fig:TwoPacket}, when the learning node predicts the pattern of its hidden neighbors only and transmits accordingly.
CPM messages are larger than CAMs with size of 500 Bytes, and the combined transmissions of CAMs and CPMs generate an average channel load of 52.1\%. 

However the reception performance improvement due to learning and predicting is less than the case with only CAMs. This is because unlike CAMs, CPMs are triggered randomly and 5 packets are emitted in a burst, making it difficult to predict the first packet of the burst. The prediction error degrades scheduling performance, affecting the packet reception ratio. Nonetheless, the PRR performance improvement via learning is 5\% at 100m and 18\% at 200m respectively.

Lastly, Figure \ref{fig:MultiPacket} shows the PRR, when the nodes transmit 750 Bytes LDM packets along with CAMs and CPMs, producing a higher average channel load of around 66.9\%. However, as the channel load increases, the performance improvement due to learning is lesser compared to the previous scenarios of lower channel loads. %with only CAMs and CAMs with CPMs. 
At higher channel loads, the learning node cannot find sufficient vacant windows of low channel activity to schedule its own packets. %Moreover, there is always a randomness in the size of the vacant window due to the random CSMA back-off windows. As neighbors transmit more,  
Nevertheless, at high channel loads, transmit rate control of DCC is supposed to be activated to prevent channel saturation, which has not been considered in this work. As part of our future work we are investigating the behavior of the learning node along with transmit rate control at high channel loads.

%% file: sections/related.tex
\section{Related work}
\label{related}

%\todo[inline]{Irfan writes related work from his field and francois too}

%\textcolor{red}{Plan of related work}
%First: related work from irfan:
%\\ - packet scheduling state of the art
%\\ - packet scheduling standard
%\\ - scheduling method that we use to compare our algorithm

%Over the years, a variety of avenues have been explored for V2X channel congestion control. The most common approach has been to by each node to periodically monitor the channel load and limit individual temporal channel resource usage i.e. transmit rate of each node \cite{limeric}, \cite{pulsar} or individual spatial channel usage i.e. transmit power \cite{dice}, \cite{fair2009}. Similarly, the work in \cite{intern} aims to control transmit rate by additionally considering application requirements. Other approaches try to enhance channel usage efficiency by optimizing the data rate \cite{miguel6mpbs} or influence the channel load monitoring by tweaking the carrier sense threshold \cite{stanica2011physical}. 

%The main essence of DCC is the decentralized measurement and control by each node, without direct coordination among nodes. However some works have suggested indirect coordination to improve a node's intelligence. The work in \cite{pulsar} proposed to extend a node's vision of the spatial channel usage via piggybacking channel load measurement via multi-hope communication. Some other approaches \cite{rezaei2007reducing} maintains an estimator of each neighbor to reduce the need for communication of state information thereby decrease congestion.

\subsection{Medium Access Control for V2X Communication}

Over the years a plethora of medium access control protocols for vehicular communication have been proposed in literature. It can be broadly categorized as contention based and contention free \cite{booysen2011survey}. Contention based algorithms involve Carrier Sense Multiple Access (CSMA), random back-off and retransmission, while contention free MAC protocols involve Time Division Multiple Access (TDMA) \cite{sahoo2013congestion,omar2013vemac} by dividing the transmission into different time frames and slots, and allocating each node different slots.

%Some works on the other hand propose a hybrid of both contention based and contention free MAC protocols \cite{su2007clustering}. Booysen et al. \cite{booysen2012performance} compare the two MAC approaches, contention based and contention free, discuss the limitations and introduce an approach called medium access with memory bifurcation. The work in \cite{djukic2009soft} proposes a software based TDMA over legacy 802.11 hardware.

One improvement of TDMA is Self-Organizing TDMA (S-TDMA) \cite{bilstrup2009ability}, where unlike centralized slot allocation of TDMA, the nodes allocate slots among themselves in a decentralized manner. Another MAC improvements is via Space Division Multiple Access or clustering nodes in geographic proximity, to handle mobility, limit channel contention, and implement spatial reuse of channel resource. The goal is to reduce interference among hidden nodes by allocating same slots to nodes sufficiently far apart \cite{javed2014joint}.
%,where a road segment is divided via SDMA and CSMA is used within each segment.

Most of these aforementioned works have intended to optimize the MAC layer scheduling for a single type of packet, mainly single hop periodic broadcast of CAM/BSM, using a fixed packet frequency, packet size and traffic pattern. However, is future there will be heterogeneity of network traffic pattern. For example a highly autonomous vehicle will communicate more compared to a human driven vehicle. Some works have analyzed multiple packet types considering strict 802.11 EDCA priority \cite{barradi2010establishing}. However other works have found the limitations of MAC layer EDCA prioritization, in the ETSI ITS stack during scarce channel resource \cite{gunther2016collective,EURECOM+5635}. 

In this work, our goal is not to introduce a new MAC protocol. Based on the standardized ITS-G5 MAC, we consider the scheduling of %single and 
multiple packet generation at the higher layers in order to generate and transmit packets to increase the reception probability by the neighbors. We propose a novel approach to reduce collision and improve packet reception performance, by increasing a node's awareness of %temporal and spatial 
channel usage via machine learning. 

\subsection{Machine Learning for V2X Communication}

Recently machine learning is being implemented for predicting various aspects of vehicular networking such as node mobility, network connectivity, network congestion control, wireless resource management etc. Ide et al. \cite{ide2015lte} uses Poisson regression trees to predict LTE network connectivity and vehicular traffic. The work in \cite{he2018integrated} uses deep reinforced learning to jointly optimize network resource allocation, caching and edge computing. In the domain of network congestion control, the work in \cite{taherkhani2016centralized} presents a centralized controller to manage channel congestion at urban intersections using k-means clustering.

A survey of machine learning for vehicular network is presented in \cite{ye2017machine}. %The survey demonstrates that ML is a promising way to take advantage of the high number of data available in future vehicular networks. 
The survey highlights the challenge of adapting the existing ML methods to these new type of networks that are highly dynamic. Besides, the survey indicates the use of RNN with LSTM as on open issue to be solved, and in this paper we propose a machine learning approach using RNN with LSTM.

%while \cite{lv2015traffic} presents an approach of predicting vehicular traffic flow using a stacked auto-encoder. 
%Xu et al. \cite{xu2014fuzzy} presents a Q-learning based handoff mechanism for heterogeneous vehicular networks. 

Moreover, existing machine learning approaches for vehicular networking do not consider a fully decentralized ad-hoc network, which we analyze in this paper. Lastly, most road safety related communications in vehicular networks involve broadcast packets, which has not been sufficiently addressed in existing studies. %In this paper we deal with single-hop broadcast network traffic, using standardized messages for safety related vehicular communication. We model the issue of learning and predicting network traffic as a time series prediction using RNN with LSTM.  

%Different ML methods have already been used for intelligent wireless resource management. \cite{ye2018deep}

%% file: sections/conclusion.tex
\section{Conclusion}
\label{conclusion}

%\todo[inline]{Irfan writes this}

In this paper we show that using recurrent neural network, a vehicle can learn and predict the transmit pattern of its neighbors. This learning can be used to schedule its own transmissions during periods of low channel activity, leading to improved packet reception ratio. In particular, most of the collisions are due to hidden nodes, therefore, learning about hidden nodes' transmit pattern and predicting hidden transmissions provides the maximum improvement. Simulation based evaluation on a dense highway scenario shows that by transmitting during periods of low channel activity, a node can significantly decrease packet collision and improve the communication performance of multiple standardized V2X safety applications on top of CSMA based channel access. %To the best of our knowledge, this is the first work to use machine learning for improving the communication performance of a decentralized CSMA based vehicular network.

%Although we demonstrated the potential machine learning using a single intelligent node,
There are still few open challenges and further work needs to be done. In a scenario with multiple learning nodes, the intelligence of the each learning node has to be coordinated with multiple learning neighbors in a decentralized manner. Similarly, the global performance in a hybrid scenario, with a varying percentage of learning nodes, i.e. some nodes having learning capability, while other nodes do not, has to be investigated. Last but not the least, transmit rate control has to be incorporated with learning, in order to  analyze the performance during high channel loads. We are currently investigating these aspects, as part of our future work.